\documentclass[]{aa}  

\usepackage{graphicx}
\usepackage{txfonts}
\usepackage{natbib}
\usepackage{amstext}

\begin{document}

\title{New powerful outburst of the unusual young star V1318 Cyg S 
(LkH$\alpha$~225)}

\titlerunning{New outburst of V1318 Cyg S}
 
   \author{T.Yu. Magakian
                    \and
          T.A. Movsessian
                    \and 
          H.R. Andreasyan
                    \and
          M.H. Gevorgyan
         }

   \institute{Byurakan Observatory NAS Armenia, Byurakan, Aragatsotn prov., 0213, Armenia\\
              \email{tigmag@sci.am}
             }

   \date{Received ...; accepted ...}

 
  \abstract
   {}
   {Young double star V1318~Cyg, which is associated with a small isolated star-forming region around HAeBe star  BD+40$^{\circ}$4124,  has very unusual photometric and spectral behavior. We present results of photometric and spectroscopic observations in the optical range.
}
   {We carried out BVRI CCD photometric observations of V1318~Cyg from 2015 Sept. to 2017 July. For the same period we acquired medium- and low-resolution spectra. Observations were performed with the 2.6 m telescope of the Byurakan observatory. We also analyzed the images of this field in IPHAS  and other surveys. 
}
   {We analyze the historical light curve for V1318~Cyg  and demonstrate that the southern component, V1318~Cyg~S, after being rather bright  in the 1970s (V$\sim$14 mag) started to lower its brightness and in 1990 became practically invisible in the optical. After its reappearance in the second half of the 1990s the star started to become very slowly brighter. Between 2006 and 2010 V1318~Cyg~S started brightening more quickly, and 
in 2015  had become brighter by more than five magnitudes in visible light. Since this time  V1318~Cyg~S has remained at this maximum. Its spectrum shows little variability and consists of a mixture of emission and absorption lines, which has allowed for estimates of its spectral type as early Ae, with obvious evidence of matter outflow. We derive its current A$_{V} \approx 7.2$ and L = 750 L$_{\sun}$ thus confirming that V1318~Cyg~S should belong to the Herbig Ae stars, making it, along with BD+40$^{\circ}$4124 and V1686~Cyg, the third luminous young star in the group. It is very probable that we observe V1318~Cyg~S near the pole and that the inclination of its dense and slow ($\approx 100$ km/s) outflow is low.
}
   {The unusual variability and other features of V1318~Cyg~S make it difficult to classify this star among known types of eruptive young stars. It could be an extreme, higher-mass example of an EXor, or an object of intermediate class between EXors and FUors, like V1647~Ori. 
}


   \maketitle
%

\section{Introduction}

\begin{figure}[h!]
  \centering
  \includegraphics[width=0.5\textwidth]{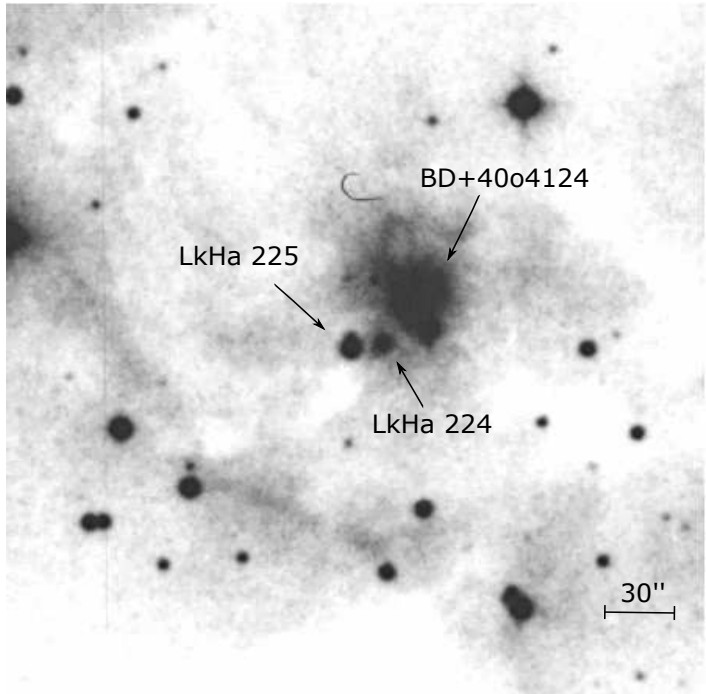}
  \caption{Direct image of the BD+40$^{\circ}$4124 region in POSS-1 
   red plate (14.06.1953), taken from SuperCOSMOS sky survey.}
   \label{field}
\end{figure}

In 1960 George Herbig published his classic paper about the Ae/Be stars connected with nebulae (lately designated as HAeBe stars) \citep{Herbig}, which were suggested to represent the more massive and luminous young stellar objects in comparison to T Tauri stars. Among them he listed the bright \object{BD+40$^{\circ}$4124} star, surrounded by several stellar objects with  fainter emission lines. One of them was designated as \object{LkH$\alpha$~225}. Only H$\alpha$ emission lines of moderate intensity were found in its medium-resolution spectrum. A general view of this group is presented in Fig.~\ref{field}.

This star attracted further attention when significant variability was found \citep{Romano,Romano1969,Wenzel}. As a variable star it received \object{V1318~ Cyg} designation. At the same time, the area around BD+40$^{\circ}$4124 was recognized as a star-forming region, where several bright IR sources were discovered. Among them, LkH$\alpha$~225 was also listed \citep{Strom1972}. This object demonstrated prominent variability even in K band \citep{Allen}.      

\begin{figure}[h!]
  \centering
  \includegraphics[width=0.5\textwidth]{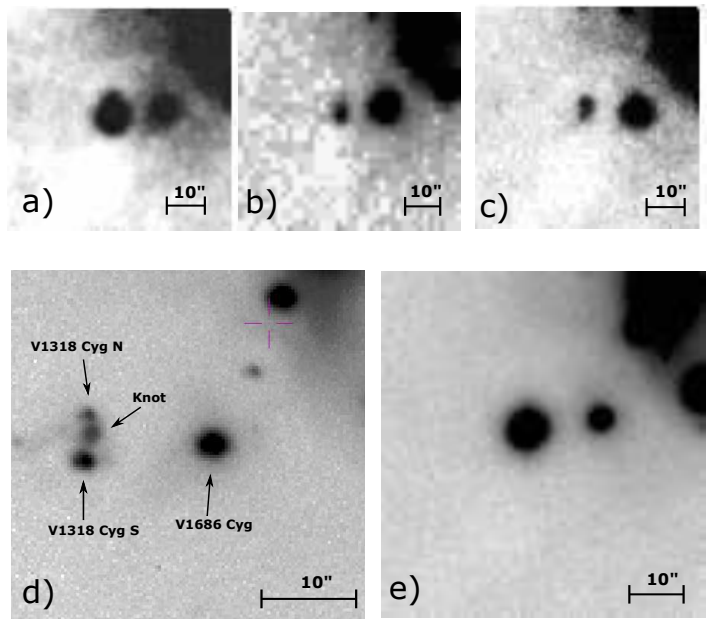}
  \caption{Enlarged direct images of V1318 Cyg field. a) POSS-1 
   red plate (14.06.1953), SuperCOSMOS sky survey; b) Quick-V plate (3.09.1983),    STScI digitized sky survey; c) POSS-2 red plate (29.06.1992), SuperCOSMOS    sky survey; d) IPHAS survey image in R band (11.10.2006), objects discussed here are marked by arrows; and e) ZTA-2.6 image in R band (22.09.2015). BD+40$^{\circ}$4124 is hidden under the bright reflection nebula in the  upper-right corner of all images.}
   \label{changes}
\end{figure}

   
It is nearly impossible to draw certain conclusions about the photometric behavior of V1318~Cyg from early brightness estimations, as was pointed out by \citet{Wenzel1980}. Usually the object demonstrated erratic variations around $V=17.0$, with occasional brightening up to 2$^{m}$ \citep{Wenzel}. One cannot exclude the longer period of high brightness in the 1950s, as such
a level can be seen on POSS-1 plate (1953) (Fig.~\ref{changes}a) and on the plate shown in \citet{Herbig} obtained in 1956. In both cases LkH$\alpha$~225 is equal in brightness to or brighter than nearby \object{LkH$\alpha$~224}. 

In the 1980s, studies of the  variability of V1318 Cyg continued in a more organized way, including systematic electrophotometric observations \citep{Ibragimov, Shevchenko1991, Shevchenko1993}. It was shown that the star fluctuated near a certain mean brightness level, sometimes demonstrating deep minima. The mean level itself was slowly decreasing from year to year,  and in 1987-1988 the star brightness was always lower than $V=17$ \citep{Shevchenko1991}.

Even in the DSS-1 images one can see that the shape of LkH$\alpha$~225 is slightly elongated in the north-south (N--S) direction (Fig.~\ref{changes}a). However, the probable duplicity of this object was mentioned for the first time only by \citet{Ibragimov}. After the deep decline in brightness, the study of the inner structure of LkH$\alpha$~225 became easier. As was shown by \citet{Aspin}, the object actually consists of two stars -- \object{V1318~Cyg~N} and \object{V1318~Cyg~S} -- and a nebulous knot between them, all placed in the N--S direction, with a total separation of 5\arcsec. This discovery immediately raised a question about the role of both components in the high-amplitude variations of visible brightness. Below we consider this point in detail.

Spectral data presented by \citet{Hillenbrand} and \citet{Magakian-a} confirmed that both objects belong to the young stars of A-F spectral type with specific emissions, including prominent forbidden lines produced by Herbig-Haro type outflow, which was found by \citet{Magakian-a}. Pronounced differences between the spectra of V1318 Cyg S in maximal and minimal brightness levels were pointed out in the same study.

\begin{table*}
\caption{Log of the V1318 Cyg spectral observations}       
\label{spectra}     
\centering        
\begin{tabular}{c c c c c}          
\hline\hline                        
Date & Spectral range (in \AA) & Resolution ($\lambda/\Delta\lambda$) & Total exposure (min) & Label
\\    
\hline                     
22.09.2015 & 5880--6740 & 2500 & 60 & A \\   
18.11.2015 & 5785--7315 & 1500 & 60 & B \\
22.11.2015 & 5785--7315 & 1500 & 60 & C \\
16.05.2016 & 5880--6880 & 2500 & 30 & D \\
10.06.2016 & 5890--6795 & 2500 & 40 & E \\
08.08.2016 & 5780--7300 & 1500 & 30 & F \\
23.08.2016 & 4120--6810 & 800 & 60 & G \\
24.08.2016 & 5895--6795 & 2500 & 40 & H \\
29.08.2016 & 4070--7055 & 800 & 20 & I \\
30.08.2016 & 5870--6875 & 2500 & 40 & J \\
06.11.2016 & 5695--7360 & 1500 & 45 & K \\
28.11.2016 & 4025--6995 & 800 & 30 & L \\
20.12.2016 & 4025--6995 & 800 & 45 & M \\
21.12.2016 & 5850--6850 & 2500 & 30 & N \\
\hline                                       
\end{tabular}
\tablefoot{Spectra A--H were acquired with TK SI-003A CCD matrix, and spectra I--N with E2V CCD42-40 matrix}.
\end{table*}

\section{Observations and data reduction}

We restarted our observations of V1318~Cyg in 2015, after the end of the refurbishing works on the 2.6 m telescope at the Byurakan observatory. The SCORPIO spectral camera in the prime focus \citep{Scorpio} was used to obtain both direct images and long-slit spectra. It was equipped with TK SI-003A $1044 \times 1044$ CCD matrix and after 2016 August with E2V CCD42-40 $2080 \times 2048$ CCD matrix.

The field of view in the imagery mode was 11.5\arcmin \ (with 0.68\arcsec/px) and 13\arcmin\ (with 0.38\arcsec/px) with TK and EEV detectors, respectively.
Direct CCD images were obtained in \textit{BVRI} bands, usually with two or three exposures for each color. The seeing was 2.5--3\arcsec. 

In the long-slit mode the width of the slit was 1.5\arcsec\ and the length was about 5\arcmin. As dispersive elements, the grism with 600 g/mm as well as  two grisms equipped with volume phase holographic gratings with 1200 and 1800 g/mm were used. For the calibration of wavelengths the comparison spectra of Ne+Ar lamp were used. In the time period from 2015 Sept. to 2016 Dec. fourteen spectra were obtained. A full log of the spectral observations is presented in Table~\ref{spectra}. Exposure times were selected to provide an S/N ratio of about 50-100 in the final spectra  after the  processing and optimal extraction.  

The direct images were processed in the standard way; after that the aperture photometry of stars was performed with the aid of the ESO-MIDAS system. For the data calibration we used  several stars in the field, covering various magnitude ranges to avoid the effects of saturation. Stars were selected from the papers of \citet{Hillenbrand} and  \citet{Shevchenko1991}, and in some cases also the magnitudes from USNO-B1 catalogue were used. A special comparison  confirmed that the magnitudes of  calibration stars had  good internal consistency. For each date the brightness of the objects was estimated separately for all exposures, after which results were averaged to obtain mean magnitudes. Typical internal scatter of results is 0\fm005--0\fm01, and typical errors of final estimates  are about 0\fm02--0\fm03. In several cases, especially for I band, we excluded from the calibration  partially saturated images of standard stars; this raised errors up to 0\fm07--0\fm08.  

  The processing of spectral data was done with the ESO-MIDAS system, following the classical procedures of long-slit spectrophotometry. All data in the paper are presented either in arbitrary flux units or are normalized to the continuum.

\section{Results}

\subsection {Photometric history and the new outburst of V1318~Cyg~}

As mentioned above, at the end of the 1980s the total visible brightness of V1318~Cyg started descending and at the beginning of the 1990s its visible brightness  fell rapidly. This fading can be seen very well in Fig.14 of \citet {Herbst}.   The question regarding which of the stars in the pair is responsible for such high-amplitude variations  can be solved by comparison of the images of the object  on DSS-1 and DSS-2 plates  (see Figs.~\ref{changes}a and~\ref{changes}c), and the data of  \citet{Aspin}. Obvious differences in morphology even on half-resolved survey images point to the significant variability of V1318~Cyg~S. Its low brightness in the 1990s, confirmed by observations of \citet{Hillenbrand} and \citet{Magakian-a},  allowed for photometric and spectral observations to be performed for both stars separately. In particular, the brightness of  both stars   was estimated by \citet{Hillenbrand} as $V=18.58, R_c=16.88, I_c=15.41$ (V1318~Cyg~N) and $V$=19.17:, $R_c$=17.34:, $I_c$=15.68: (V1318~Cyg~S). This corresponds to an integral brightness of the double star of about $V=18.1$. Such a value is in good compliance with the measurements of \citet{Shevchenko1991} for 1987-1988, which in turn means that by the end of the 1980s V1318~Cyg~S has possibly reached a minimal level of brightness. 

One cannot exclude the variability of V1318~Cyg~N however, and indeed the extreme faintness of V1318~Cyg in the falls of 1995 ($V\approx21$) and 1997 ($V\approx24$), according to the estimates of \citet{Herbst} is evidence supporting this statement.

Beginning new observations of V1318~Cyg field in 2015 September, we immediately found that  the star underwent a new powerful outburst and its brightness reached the level of the 1960s or even higher (Fig.~\ref{changes}e). The position of the photometric peak of the image was pointing to V1318~Cyg~S as the outbursted star. The faint northern star was lost in its glare
in our very short (to avoid saturation effects) exposures. 

To find precise coordinates of both stars we consulted the Gaia DR2 catalogue \citep{gaia2016,gaia2018}, where both V1318 Cyg N and  V1318 Cyg S stars are resolved and listed (DR1 did not include these objects). We performed the astrometric calibration of our images, obtained within the years of 2015-2017, with the aid of the \textit{astrometry.net }system \citep{Lang}, and measured the position of the outbursting star. All images demonstrated very good internal agreement and the dispersion of stellar position was $\approx$0.27\arcsec\ (which is about half a pixel) in both coordinates, for \textit{V} and \textit{I }colors. Our coordinates agree with the Gaia DR2 position of the V1318~Cyg~S star within 0.3\arcsec.  Gaia DR2 photometry also confirmed the outburst of the southern star: G magnitudes are 13.23 for the southern star and 18.92 for the northern star, that is, they currently differ in brightness by about 6$^{m}$. 

\begin{table}
\caption{V1318 Cyg (LkH$\alpha$ 225) photometry}       
\label{phot1}     
\centering        
\begin{tabular}{c c c c c}          
\hline\hline                        
Date & \textit{B} & \textit{V} & \textit{R} & \textit{I} \\    
\hline                     
22.09.2015 & 16.95 & 14.61 & 12.53 & 11.06 \\   
18.11.2015 & 17.18 & 14.76 & 12.72 & 11.16 \\
31.03.2016 & 17.38 & 14.98 & 12.95 & 11.20: \\
10.06.2016 & 16.78 & 14.40 & 12.43: & 10.85: \\
23.08.2016 & 16.77 & 14.38 & 12.49 & 10.96: \\
20.12.2016 & 16.81 & 14.46 & 12.59 & 10.92 \\
20.07.2017 & 17.48 & 15.03 & 13.12 & 11.29 \\
\hline                                       
\end{tabular}
\end{table}

Though we have found on our exposures the faint trace of  V1318~Cyg~N near the northern edge of the bright southern star, it was not possible to resolve both objects to measure them separately. Therefore, the photometry presented below is related to both stars and the nebula between. Photometric data are presented in Table~\ref{phot1}. Their errors do not exceed 0\fm02--0\fm03, as was already noted above; if errors are higher (up to 0\fm08), values are marked with ``:''. In any case, taking into account the Gaia magntude for V1318~Cyg~N, it is easy to estimate that its contamination affects the values in Table~\ref{phot1} no more than 0\fm01.

As can be seen from our photometric data, the star  presently remains at an almost constant level of brightness. To find an answer to the immediate question of when this new large-amplitude increase
of the brightness of V1318 Cyg S started, we checked various surveys and found the images of IPHAS survey \citep{Drew}, obtained from 2003 to 2006, where both stars are resolved and V1318 Cyg S in \textit{R} band is already brighter than V1318~Cyg~N by at least one magnitude. We performed photometry for both stars using all suitable IPHAS images. Because of the relative faintness of the stars, the influence of the central nebular knot (Fig.~\ref{changes}d) on the photometric results was significant. This fact as well as the varying quality of IPHAS exposures led to significant measurement errors, reaching several tenths of a magnitude. We tried to minimize these errors as much as possible, and present the best of our estimates in Table~\ref{phot2}.

These data, when we compare them with the measurements of \citet{Hillenbrand}, lead us to two important conclusions. On the one hand, V1318~Cyg~N became about two magnitudes fainter in comparison with 1993  Jan. This confirms the appreciable variability of the northern star. On the other hand, we see that  the brightness of V1318~Cyg~S in 2003-2006 also is lower  than in 1993  Jan  by several  tenths of a magnitude. This means that after the great minimum of the second half of the 1990s its level of brightness was restored to the values observed at the end of the 1980s, and until the end of 2006 there was no further increase.

The  V1318 Cyg field was also observed on 2003 Sept. 20 in the SDSS survey -- see Data Release 12 \citep{DR12}. We do not show this image here, because the appearance of the object does not significantly differ from IPHAS images. SDSS photometric data related to V1318~Cyg~S do not exist, since the image- recognizing program mistook the pair of stars with the middle nebula for a galaxy and integrated all their fluxes.    

\begin{table}
\caption{V1318~Cyg~S and V1318~Cyg~N brightness on the IPHAS images}    
\label{phot2}     
\centering        
\begin{tabular}{c c c}          
\hline\hline                        
Date & \textit{R} & \textit{I} \\   
\hline                 
\multicolumn{3}{c}{V1318 Cyg S} \\
\hline    
09.08.2003 & 17.66 & 16.45 \\   
10.08.2003 & 17.67 & 16.45 \\
18.10.2003 & 18.01 & 16.04 \\
08.08.2004 & 17.72 & 16.23 \\
11.10.2006 & 17.86 & 15.94 \\
\hline
\multicolumn{3}{c}{V1318 Cyg N} \\
\hline
09.08.2003 & 18.48 & 17.44 \\   
18.10.2003 & 19.17 & 17.58 \\
08.08.2004 & 18.93 & 18.02 \\
11.10.2006 & 18.79 & 17.21 \\
\hline
\end{tabular}
\end{table}

The most recent  images of the region around BD+40$^{\circ}$4124 that precede our new observations are  available in the Pan-STARRS1 data archive \citep{Chambers}. They show both the N and S stars resolved, and V1318~Cyg~S seems definitely brighter than in IPHAS images. In view of the ``stacked'' nature of DR1 images (i.e., they represent co-added images made from the multiple exposures taken over the survey),  only the mean epoch of observations for  V1318~Cyg (May 2012) was given. But while the current paper was in preparation, the new data release (DR2) of the Pan-STARRS survey became available, which contains the separate images of this area, and the dates of observations span  2010 Aug. - 2014 Oct..  Analyzing these images, we found that even in 2010 the star  V1318~Cyg~S was brighter than on SDSS images of 2003, and a further perceptible increase in its brightness took place in the middle of 2013. Photometric estimates (magnitudes in\textit{ grizy} bands)
provided by Pan-STARRS DR1 were converted to \textit{BVRI} magnitudes using the transformations suggested by \citet{Jester}, which resulted in the following values for the brightness of both stars: V1318~Cyg~S --
$B =  18.73; V = 16.38; R = 14.57; I = 12.90$; V1318~Cyg~N --
$B =  21.14; V = 19.65; R = 17.91; I = 16.30$. One can see that these $R$ and $I$ magnitudes are brighter than IPHAS values by more than 3$^{m}$  for V1318~Cyg~S and by about 1$^{m}$ for V1318~Cyg~N. On the other hand,  they are still about 2$^{m}$ lower than the present level. These estimates however should be used with some caution because as we found by comparison with other available photometric data,   highly variable emission background in the central part of BD+40$^{\circ}$4124 cloud can introduce serious systematic errors in Pan-STARRS photometry.

\begin{figure}[h!]
  \centering
  \includegraphics[width=0.50\textwidth]{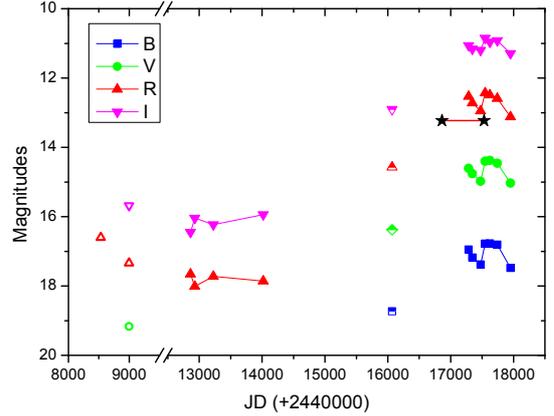}
  \caption{\textit{BVRI} light curve of V1318~Cyg~S for the period of 1991-2017. Filled symbols show our measurements for IPHAS and Byurakan images; open symbols  show data of \citet{Aspin} and \citet{Hillenbrand}; half-open symbols  show mean estimates for the four-year period of time from PanSTARRS survey (see text); and the line bounded by asterisks shows the level in G magnitude and time period of observations from \textit{Gaia} DR2.}
   \label{lightcurve}
\end{figure}

We placed all measurements of  V1318~Cyg~S on the light curve (Fig.~\ref{lightcurve}).
Taking  all results into account, we come to the following conclusions.
Somewhere between 2000 and 2003 V1318~Cyg~S came out of its minimum and returned to the mean brightness level of 1988. This star then remained at the same brightness for at least three years without pronounced changes. In the time period between 2006 and 2010 the brightness of V1318~Cyg~S started to raise to more than 5 mag (at an unknown speed), reaching a maximum by the second half of 2015,  and remaining at this level for at least two years\footnote{When this paper was ready for publication, we obtained new images of the region under study, which show that V1318~Cyg~S retains its brightness; thus, the duration of the maximum is already three years or even more.} Another long maximum possibly took place in the 1950s, but no systematic observations were performed at that time (see above). We also see that small fluctuations near the present maximum do not lead to noticeable changes of color indices of the star, in contrast with its behavior at the lower level of brightness in the 1980s \citep{Ibragimov,Herbst}. 

\subsection {Spectrum of V1318~Cyg~S at its maximum}

\subsubsection {General appearance}

\begin{figure}[h!]
  \centering
  \includegraphics[width=0.5\textwidth]{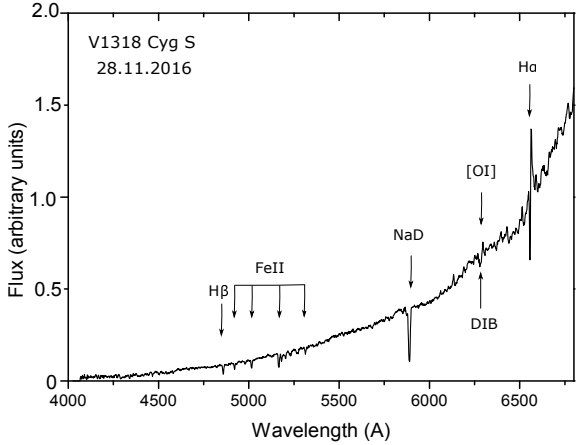}
   \caption{Low-dispersion spectrum of V1318~Cyg, obtained with the 2.6 m telescope. Main features are identified.}
   \label{genspec}
\end{figure}

\begin{figure}[h!]
  \centering
  \includegraphics[width=0.45\textwidth]{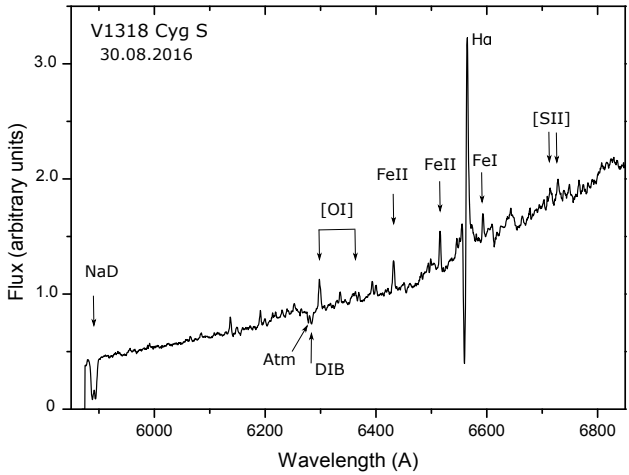}
   \caption{Red part of V1318~Cyg spectrum with better dispersion. The spectrum was obtained with the 2.6 m telescope. P Cyg profile of H$\alpha$ line and strong NaD absorptions are prominent.  Several other features, including forbidden lines,  are also identified. A part of this spectrum is shown in more detail in Fig.\ref{emlines}.}
   \label{genspec_b}
\end{figure}

All our spectrograms of V1318~Cyg at maximal brightness are remarkably similar and do not show any prominent changes during the 1.5 year time interval of observations. Furthermore, they resemble the star spectrum during its previous maximum in the 1970s, which was briefly described by \citet{Magakian-a}, but on the other hand noticeably differ from the spectrum at minimum brightness \citep{Hillenbrand}. The general appearance of the V1318~Cyg spectrum is shown in Figs.\ref{genspec} and \ref{genspec_b}.
Due to the difference of  almost six magnitudes in brightness, the influence of V1318~Cyg~N should be negligible; in any case, slit width was significantly lower than the distance between the stars. In the following discussion we consider these data as the V1318 Cyg S spectra.
 
\begin{figure}[h!]
  \centering
  \includegraphics[width=0.5\textwidth]{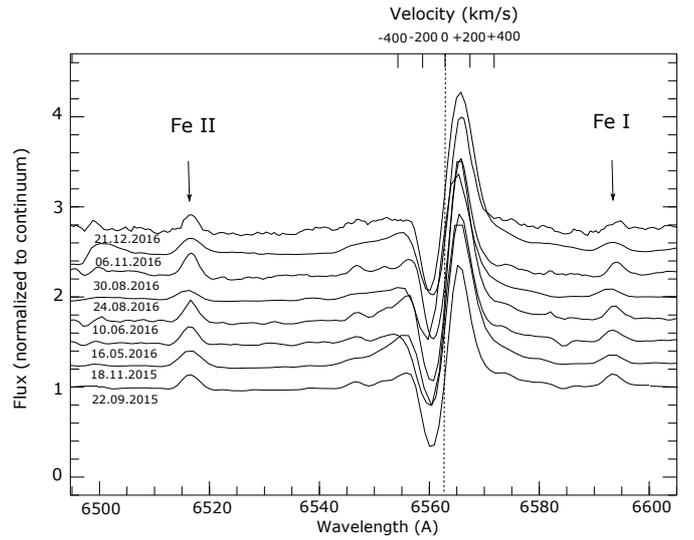}
  \caption{H$\alpha$ line in V1318~Cyg spectrum: comparison for various dates.}
   \label{halpha}
\end{figure}

One can see a very red continuum with both emission and absorption features. The only line with a complex profile is H$\alpha$.  More detail is shown in Fig.\ref{halpha} where we present H$\alpha$ profiles obtained from all our better-resolution spectra. To compensate small positional differences between spectra introduced by rebinning,  for some of them we added slight ($\le0.1\AA$) shifts to adjust all spectra by two nearby Fe emission lines. Iron emission in T Tau and HAe/Be stars usually has  low variations in radial velocity, with one being close to stellar velocity (see, e.g. \citealt{Petrovbabina}). As one can see, this profile has the typical appearance of rather broad (nearly 800 km/s in total width)  H$\alpha$ emission, divided by a blue-shifted, relatively narrow absorption component, which goes well down the continuum; we did not find any significant changes in its structure or in radial velocities, though the equivalent widths (EWs) of both red-shifted and blue-shifted emission components vary to some extent. Such double-peaked H$\alpha$ profiles are common in young active stars, but their central absorptions often do not reach the continuum level.

 Besides the the P~Cyg component of the H$\alpha$ line, we see prominent absorptions of NaD and, in the blue part of the spectrum, strong absorptions of H$\beta$, \ion{Fe}{ii} (42), and (48), \ion{Mg}{i} (2) as well as several weaker ones  (Fig.\ref{genspec}; at the shorter wavelengths weakness of continuum prevents detection of any lines except for  the trace of H$\gamma$ absorption). These lines are typical for the early A spectral type but, having negative radial velocities (see details below), they definitely belong to the outflowing envelope of the star.  In addition, we detected a prominent diffuse interstellar band (DIB) $\lambda$6284 \AA, which easily resolves from the nearby atmospheric absorption $\lambda$6278 \AA,\ and is among the strongest ones \citep{Herbig1995}. In the red part of the spectrum the only undoubtedly observed metallic absorption is $\lambda$6456 of \ion{Fe}{ii} (74); there is no convincing evidence of the existence of the famous \ion{Li}{i} $\lambda$6707 absorption.

In fact, with our resolution we did not detect any photospheric absorption lines, which makes the spectral type of V1318~Cyg~S  difficult to estimate.
This question is discussed in detail in the following section.
    
\subsubsection{Emission lines}

Virtually all emission lines in the spectrum of V1318~Cyg~S in maximal brightness, covered by our spectrograms, are located in the $\lambda\lambda$ 6000--7000 \AA\ range (see Fig. \ref{genspec_b}). They can be divided in three groups: H$\alpha$ (the only hydrogen line with an emission component), which was described above, forbidden lines of [\ion{O}{i}] and [\ion{S}{ii}], and a large number of low-excitation \ion{Fe}{i} and \ion{Fe}{ii}  lines.  

Forbidden emissions are contaminated not only by night sky lines of [\ion{O}{i}], but also by lines of the diffuse emission background. Lines belonging to the star itself are not especially prominent and are produced in the HH outflow originating from V1318~Cyg~S or perhaps from both stars in the pair \citep{Magakian-a}. It is worth mentioning that [\ion{S}{ii}] lines are wide and probably consist of two poorly resolved components, though they do not show particularly conspicuous variations as in \object{PV~Cep} (\citealp{Magakian2001}; but the time period, encompassed by present observations of V1318~Cyg~S, is also shorter).

\begin{figure}[h!]
  \centering
  \includegraphics[width=0.45\textwidth]{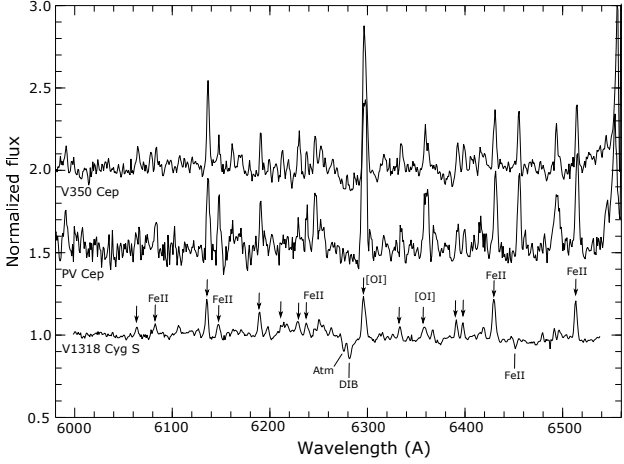}
  \caption{Emission lines in V1318~Cyg, PV~Cep and V350~Cep: comparison. All main features are identified; lines, marked only by arrows, belong to Fe I.}
   \label{emlines}
\end{figure}

Nearly all other emissions in the red range of the V1318~Cyg~S spectrum belong to \ion{iron, especially Fe}{i}, which makes it similar to T~Tau-type stars. We show this in more detail in Fig.\ref{emlines}, with matched spectra of  well-known eruptive young stars PV~Cep and \object{V350~Cep}, obtained in Byurakan with the same equipment on 24.08.2016 and 22.09.2015, respectively.
All emission features seen in this range were identified with lines of \ion{Fe}{i} and \ion{Fe}{ii} using the spectrum of \object{VY~Tau} at maximum brightness for reference \citep{Herbig1990}.

 One can see the remarkable similarity between the spectra of both active stars,  which closely resemble each other even in fine detail (though V350~Cep presumably belongs to T~Tau-type with spectral type M2, while the probable spectral type of PV~Cep is A5);  on the other hand, V1318~Cyg~S differs from its counterparts in that it has significantly lower EWs of emission lines and  obviously lower excitation level. Especially spectacular is the absence of intense emissions of $\lambda$~6456.38 \AA\ \ion{Fe}{ii}(74) (this line  is actually seen in absorption) and of the strong blend of $\lambda$ 6493.78 and $\lambda$ 6499.65\AA\ \ion{Ca}{i}(18). Beyond the range shown in Fig.\ref{emlines}, the spectra of the three stars completely differ: for example,  both V350~Cep and PV~Cep demonstrate rich emission spectra also at shorter wavelengths.

\begin{table*}
\caption{Equivalent widths of selected V1318~Cyg~S spectral lines}
\label{widths}     
\centering        
\begin{tabular}{c c l l l}          
\hline\hline                        
Line & EW, \AA & spectra\tablefootmark{1} & PV~Cep\tablefootmark{2} & V350~Cep\tablefootmark{2} \\
\hline                     
$\lambda$6731 [\ion{S}{ii}] (2F) & $-0.36\pm{0.11}$ & A-F,H,J-K & &\\
$\lambda$6717 [\ion{S}{ii}] (2F) & $-0.25\pm{0.06}$ & A-F,H,J-K & &\\
H$\alpha$ (em) & $-6.86\pm{1.50}$ & A-F,H,J-K & & \\
H$\alpha$ (abs) & $1.91\pm{0.61}$ & A-F,H,J-K & & \\
$\lambda$6517 \ion{Fe}{ii (40)} & $-0.55\pm{0.13}$ & A-F,H,J-K & $-2.34$ & $-1.23$\\
$\lambda$6456 \ion{Fe}{ii (74)} & $0.43\pm{0.14}$ & A-F,H,J-K & & \\
$\lambda$6430 \ion{Fe}{i (62)} & $-0.72\pm{0.10}$ & A-F,H,J-K & $-1.99$ & $-1.27$\\
$\lambda$6300 [\ion{O}{i}] (1F) & $-1.28\pm{0.20}$ & A-F,H,J-K & $-5.26$ & $-3.92$\\
$\lambda$6278 DIB & $0.64\pm{0.10}$ & A-F,H,J-K & & \\
$\lambda$6137 \ion{Fe}{i (207)} & $-0.69\pm{0.08}$ & A-F,H,J-K & $-1.69$ & $-1.61$\\
Na D & $8.44\pm{0.23}$ & B-C,F-G,J-M & & \\
$\lambda$5018 \ion{Fe}{ii (42)} & $2.47\pm{0.07}$ & G,L-M & & \\
$\lambda$4923 \ion{Fe}{ii (42)} & $1.89\pm{0.18}$ & G,L-M & & \\
H$\beta$ & $4.26\pm{0.38}$ & G,L-M & & \\

\hline                                       
\end{tabular}
\tablefoot{ EW values of emission lines are negative.\\
\tablefoottext{1}{Spectra, used for averaging, are indicated by letters, which correspond to the Table~\ref{spectra}.}\\ 
\tablefoottext{2}{PV~Cep was observed at 24.08.2016 and V350~Cep at 22.09.2015} }

\end{table*}

\subsubsection{Equivalent widths and radial velocities}

Estimations of EWs of selected emission and absorption lines were performed using the best 12 spectra from our data set. In Table~\ref{widths} we present their averaged values, taking into account that the spectrum of V1318~Cyg~S at maximal brightness has not significantly changed in two years. Especially noteworthy is the almost constant strength of the NaD doublet, which is the most intense absorption in the star spectrum. Significant variations of intensities (which can also be seen from the standard deviations from Table~\ref{widths})  were only detected in the components of the H$\alpha$ line: their EWs were changing in the range of 4 -- 10\AA\ (emission) and 1 -- 3\AA\ (absorption).

In general the strength of emission spectrum of V1318~Cyg~S is not high. For comparison
we present in Table~\ref{widths} the EW values for typical lines in the spectra of PV~Cep and V350~Cep, shown also in Fig.\ref{emlines}. As one can see, they vastly exceed those of V1318~Cyg~S.

   We also estimated the heliocentric radial velocities of strong and well-defined lines in the V1318~Cyg~S spectrum, separately measuring emission and  absorption components of Balmer lines, the NaD pair, the forbidden lines of [\ion{O}{i}] and [\ion{S}{ii}], the four strongest \ion{Fe}{ii} lines and $\lambda$5183 \ion{Mg}{i}, observed in absorption in the blue range, and $\lambda$6456 \ion{Fe}{ii} observed in red range, as well as selected unblended emissions of \ion{Fe}{i} ($\lambda\lambda$ 6137, 6191, 6393, 6400) and \ion{Fe}{ii} ($\lambda\lambda$ 6432, 6516). We would like to mention that for the measurements of lines in the red range we used 0.80 and 0.50 \AA/pix resolution spectra; on the other hand, we had only 1.50 \AA/pix spectra to measure H$\beta$ and metallic absorptions in the blue spectral range. Radial velocities, averaged over all available measurements, are collected in Table~\ref{vel}. Standard deviations of mean  radial velocities of hydrogen lines  are clearly larger than observational errors (which are about $15-20$ km s$^{-1}$), definitely confirming some variations in time. However, within the accuracy of our data we did not find any reliable trends in such variations. We further discuss radial velocities in the following section.   

\begin{table}
\caption{Mean heliocentric radial velocities of V1318~Cyg~S spectral lines}   \label{vel}     
\centering        
\begin{tabular}{c c c}          
\hline\hline                        
Line & V$_{R}$ km/s & dates\tablefootmark{1} \\    
\hline                     
H$\alpha$ (em) & $+107\pm{30}$ & A,D,E,H,J,K,N \\
H$\alpha$ (abs) & $-150\pm{29}$ & A,D,E,H,J,K,N \\
Na D (abs) & $-81\pm{16}$ & J,K,N \\
H$\beta$ (abs) & $-90\pm{41}$ & I,L,M \\   
\ion{Fe}{ii},\ion{Mg}{i} (abs, blue range) & $-45\pm{23}$ & I,L,M \\
\ion{Fe}{ii} (abs, red range)\tablefootmark{2} & $-68\pm{32}$ & A,E,H,J,N \\
\ion{Fe}{i} (em) & $-14\pm{17}$ & A,E,H,J,N \\
\ion{Fe}{ii} (em) & $-21\pm{22}$ & A,E,H,J,N \\
$\lambda$6300 [\ion{O}{i}] (em) & $-89\pm{22}$ & A,D,E,H,J,K,N \\
\ion{[S}{ii}] (em) & $-127\pm{20}$ & A,D,E,H,J \\ 
\hline                                       
\end{tabular}
\tablefoot{
\tablefoottext{1}{Spectra used for measurement and averaging are indicated by letters, which are the same as in Table~\ref{spectra}.}\\
\tablefoottext{2}{Only $\lambda$6456 \ion{Fe}{ii (74)} line. }
}
\end{table} 

\section{Discussion and conclusions}
     
Starting from the paper of \citet{Aspin}, V1318~Cyg~S was considered a luminous (L$_{bol} \approx\ $1600 L$_{\sun}$) and strongly reddened object. However, all these estimates are based on the studies in infrared and sub-millimeter range \citep{Palla,vandenAncker}. On the other hand,  spectral classification of HAeBe stars may encounter problems,  which can lead to serious discrepancies in spectral types \citep[see][for detailed discussion]{Hernandez}. 

Comparing the optical spectra of V1318~Cyg~S during the period of maximal brightness with spectral atlases and   surveys of HAeBe stars, we found that in the blue spectral range \ion{well noticeable Fe}{ii} (42) absorptions and several other features make its spectrum  remarkably similar to that of \object{XY~Per} Herbig Ae star (A5), shown in the atlas of \citet{Gray}, as well as to such HAeBe stars as \object{UX~Ori (A3)} \citep{Hernandez} and \object{RR~Tau} (A0) near the maximum \citep{CK,Rodgers} (spectral classes are given according to work of \citealt{Hernandez}). On the other hand, typical HAeBe stars have more prominent broad absorption wings of Balmer lines, and certain details, especially the powerful NaD absorptions, seem to contradict the early spectral type. However, such  ``anomalous'' \citep{Hernandez} NaD and  \ion{Fe}{ii} absorption features in some HAeBe stars have a non-photospheric origin,  
being caused by material surrounding the star. In the case of V1318~Cyg~S, high negative radial velocities of these lines (see Table~\ref{vel}) confirm their formation in the envelope or in the dense stellar wind. 

Assuming that V1318~Cyg~S
indeed belongs to HAeBe stars, it should be noted that given the strength of spectral lines it is not similar to the majority of these stars. Comparing equivalent widths from Table~\ref{widths} and corresponding values from Table 8 from the study of \citet{Hernandez}, one can see that the greatest resemblance (both in strength of \ion{Fe}{ii} absorptions and weakness of H$\alpha$ emission) to V1318~Cyg~S demonstrate XY~Per, UX~Ori, RR~Tau, \object{BF~Ori} and \object{LkH$\alpha$~208} stars. As one can see, all these stars belong to the group of so-called UX~Ori-type variable stars (UXors). We return to this topic below.

On the other hand, emission lines in the red part of the spectral range and strong NaD absorption make V1318~Cyg~S somewhat similar to T Tau-type stars, though there is no evidence for a later-type spectrum, at least at maximal brightness. In fact, the observed combination of Fe absorption in the blue spectral range and Fe emissions in the red is quite unusual. 
The key to the problem could be the difference between their radial velocities
(Table~\ref{vel}). Taking into account the heliocentric velocity of the BD+40$^{\circ}$4124 molecular cloud \citep[$-$9 km/s;][]{Loren}, one can see that low radial velocities of Fe emissions (which very probably originate near the  stellar surface) are close to this value.
In fact, it is reasonable to consider their mean value of $-18$ km/s as the systemic radial velocity of V1318~Cyg~S. Consequently, the higher negative values from Table~\ref{vel}  will correspond to the speed of various layers of the expanding envelope and collimated outflow.

The existence and features of the matter outflow near V1318~Cyg~S were studied in detail in the optical \citep{Magakian-a}, the infrared, the sub-mm \citep{vandenAncker, Davies, Looney, Sandell}, and in radio \citep{Palla,Matthews}. All studies of CO emission confirm the existence of a molecular outflow, oriented close to the line of sight, though there is ambiguity about its possible source: it could be V1318~Cyg~S \citep{Matthews} or could be heavily embedded and non-detectable in the optical range protostar located nearby \citep{Looney}. In any case, the presence of  high-velocity wind from V1318~Cyg~S itself is obvious in view of many shock-excited spectral lines detected in optical \citep{Magakian-a} and infrared \citep{vandenAncker} ranges.
One can assume that
V1318~Cyg~S, like a great number of HAeBe and T Tau stars, is driving a more or less collimated outflow, creating Herbig-Haro-type emission in its vicinities. Indeed, forbidden emissions in the nebula between both stars have radial velocities of about $-$80 km/s  \citep{Magakian-a}, which agrees well with results from Table~\ref{vel} (one should keep in mind that other measurements from this latter paper refer in fact to V1318~Cyg~N, because the southern star was very faint during that period).
Absence of \ion{[N}{ii]} lines and moderate equivalent  widths of outflow emissions point to the low excitation level, corresponding to a shock velocity about 100 km s$^{-1}$ or lower.   Low inclination of the V1318~Cyg~S flow to the line of sight is also corroborated by the shape of stellar H$\alpha$ line profile: as is shown by \citet{Kurosawa}, such kinds of profiles \citep[Type III-B by the classification of][]{Reipurth} require slow accelerating wind viewed near pole-on. The significant equivalent width of blue-shifted absorption lines, which presumably arise in dense wind or/and envelope, should also be taken into account.

Independent evidence for a near pole-on orientation of V1318~Cyg~S
is provided by a high-resolution polarization map obtained by \citet{Perrin}, showing low polarization at the position of V1318~Cyg~S and the centrosymmetric pattern of polarization vectors around it, without any traces of a circumstellar dust disk.

To understand the main parameters of the star we should first of all find the amount of interstellar extinction toward V1318~Cyg~S as well as toward two other more massive members of this small association, namely BD+40$^{\circ}$4124 and \object{V1686~Cyg} (LkH$\alpha$~224). However, up to now almost all such estimates have been indirect and uncertain. For example, analysis of the molecular emissions in the infrared and radio ranges resulted in improbably high values of A$_{V}$ of 25 \citep{Aspin} and even 50 \citep{Palla}, which contradict the direct visibility of V1318~Cyg~S in the optical range.  The estimation of A$_{V} = 10$, based on the approximation of the spectral energy distribution  \citep{Aspin} is more reliable. In the same way, \citet{Hillenbrand} estimated A$_{V} > 8$ and A$_{V} > 7$ for V1318~Cyg~S and V1318~Cyg~N, respectively. However, in the same work the extinction values for the brightest stars were calculated in a straightforward way on the basis of established spectral types and previously observed $(B-V)$ colors. These latter authors found A$_{V} = 3.6-3.8$ for BD+40$^{\circ}$4124 and 4.2--6.7 for V1686~Cyg, that is, much lower values. In the same way, the value of A$_{V} = 3.0$ for these stars was obtained in the comprehensive paper of \citet{vandenAncker}; but for V1318~Cyg~S they get A$_{V} = 15.4$, again based on the analysis of the features in the infrared spectrum.   

Presently, while V1318~Cyg~S remains at a stable maximum, its spectral type could be more or less reliably assigned as early A. Furthermore, since the current colors are well determined, we can directly derive A$_{V}$. We computed mean $(B-V)$ and $(R-I)$ from Table~\ref{phot1} and compared them with intrinsic colors of A2V and A5V stars from the tables of \citet{Johnson1963,Johnson1966}, following the procedure and corrections described by \citet{Hillenbrand}. As a result we derived A$_{V}$ = 7.22 (for A2V star) or 7.00 (for A5V star) using $(B-V)$, and 7.57 or 7.38, respectively, using $(R-I)$. The excellent agreement of these values however contradicts the larger estimates listed above. It is possible to assume that after prominent brightening, some amount of absorbing dust was cleared from the line of view. Thus, it will be interesting to repeat the observations of the object in the infrared range and to compare them with previous data.

\textit{Gaia} project DR2 allows for reliable estimates to be made of the V1318~Cyg group distance. Indeed, one should assume that at least BD+40$^{\circ}$4124, the brightest star in this small cluster, V1686~Cyg,  and both stars of V1318~Cyg are close neighbors in space. The trigonometric parallax of BD+40$^{\circ}$4124, taken from \textit{Gaia} DR2, is $1.092\pm0.031$ mas, which corresponds to a distance of 916 pc. V1686~Cyg,  as well as  V1318~Cyg~S and V1318~Cyg~N, were 
also detected and measured by \textit{Gaia}, but with errors one order higher, probably due to their relative faintness. For completeness  we list these data here: $0.785\pm0.122$, $1.393\pm0.292,$ and $1.490\pm0.735$  mas, respectively. It seems reasonable to use the value of 920 pc for the distance of the group, which is also in excellent agreement with the value of 980 pc estimated by \citet{Shevchenko1991} and used by other researchers. With $\overline{V} = 14.66, A_{V} = 7.2$, and this new distance, one can derive for V1318~Cyg~S at maximum M$_{V} = -2.36$ and L = 750 L$_{\sun}$ in the optical range. This value, though naturally higher, agrees reasonably well  with 590 L$_{\sun}$ estimated by \citet{vandenAncker}, and confirms that V1318~Cyg~S should be a sufficiently massive YSO. The value of bolometric luminosity is of course at least two times larger according to the previous estimates. 

It is not easy to attribute V1318 Cyg S to one of the known types of erupting medium-mass young stars. Let us discuss the existing options.

\textit{FUor-type star}. FUors and FUor-like stars are very rare. Their specific features are now well-defined, being described in recent reviews \citep{Audard,Connelley}. The star V1318~Cyg~S cannot be considered as a FUor, or even an unusual one, because its spectrum in both optical and infrared range is not similar to typical FUors; besides, it is definitely a more massive object.

\textit{EXor-type star}. EXors are usually considered to be lower-scale versions of FUors, but this class is loosely defined and probably not homogeneous \citep{Audard,Lorenzetti}.
The light curve of V1318~Cyg~S bears a resemblance to those of some EXors; another young eruptive star with developed FeI emission spectrum at maximum brightness is VY Tau \citep{Herbig1990}, also an EXor. However, the high luminosity of V1318~Cyg~S rules out such a classification: the luminosity of all known and probable EXors is lower by at least in one order of magnitude even in outburst.

\textit{Intermediate (V1647 Ori-like) object}. This is a poorly defined and yet-to-be properly characterized class of young eruptive objects. However, they have some common features, namely they are connected with small nebulae and collimated outflows (like FUors); show outbursts with durations of several years (in between FUors and EXors); have T~Tau-type spectra (like EXors); and their bolometric luminosities are moderate. In fact, V1318~Cyg~S satisfies the first two criteria, and its emission spectrum shows a partial similarity to T~Tau type. However, the bolometric luminosity of V1318~Cyg~S cannot be considered as moderate. It could be a larger-mass member of this class (if such a class indeed exists); one should keep this option in mind.

\textit{HAeBe/UXor-type variability}. We mentioned above that the intensity of the main emission and absorption lines in the V1318~Cyg~S spectrum resemble those of known UXors. However, many arguments suggest that V1318~Cyg~S is observed close to polar regions; thus, UXor-type occultations by clumps, moving close to the equatorial plane, are impossible. Naturally, one cannot exclude 
a role of another kind obscuration event  in the unusual photometric behavior of  V1318~Cyg~S.

In any case, when considering possible explanations for the variability of V1318~Cyg~S, one should take into account such features as a very large brightness amplitude in the optical range (more than 5 mag, possibly even more than 7 mag), large amplitude variations even in near-infrared, very long duration minimum (10 years or more), and pronounced spectral changes between minimum and maximum. The duration of the recent outburst, which is still in progress, should
also be regarded as unusual, comparing it to previous, though incomplete photometrical data for this star.

We hope that continued observations, especially collecting and analyzing high-resolution optical spectroscopic data, will shed more light on the nature of this star.

\begin{acknowledgements}
The authors are grateful to referees for important advice. This work was supported by the RA MES State Committee of Science, in the frames of the research projects number 15T-1C176 and 18T-1C-329.

This paper makes use of data obtained as part of the INT Photometric H$\alpha$ Survey of the Northern Galactic Plane (IPHAS) carried out at the Isaac Newton Telescope (INT). The INT is operated on the island of La Palma by the Isaac Newton Group in the Spanish Observatorio del Roque de los Muchachos of the Instituto de Astrofisica de Canarias. All IPHAS data are processed by the Cambridge Astronomical Survey Unit, at the Institute of Astronomy in Cambridge. The bandmerged DR2 catalogue was assembled at the Centre for Astrophysics Research, University of Hertfordshire, supported by STFC grant ST/J001333/1.

SDSS is managed by the Astrophysical Research Consortium for the Participating Institutions of the SDSS Collaboration.
The Pan-STARRS1 Surveys (PS1) and the PS1 public science archive have been made possible through contributions by the Institute for Astronomy, the University of Hawaii, the Pan-STARRS Project Office, the Max-Planck Society and its participating institutes, the Max Planck Institute for Astronomy, Heidelberg and the Max Planck Institute for Extraterrestrial Physics, Garching, The Johns Hopkins University, Durham University, the University of Edinburgh, the Queen's University Belfast, the Harvard-Smithsonian Center for Astrophysics, the Las Cumbres Observatory Global Telescope Network Incorporated, the National Central University of Taiwan, the Space Telescope Science Institute, the National Aeronautics and Space Administration under Grant No. NNX08AR22G issued through the Planetary Science Division of the NASA Science Mission Directorate, the National Science Foundation Grant No. AST-1238877, the University of Maryland, Eotvos Lorand University (ELTE), the Los Alamos National Laboratory, and the Gordon and Betty Moore Foundation.
This work has made use of data from the European Space Agency (ESA) mission
{\it Gaia} (\url{https://www.cosmos.esa.int/gaia}), processed by the {\it Gaia}
Data Processing and Analysis Consortium (DPAC,
\url{https://www.cosmos.esa.int/web/gaia/dpac/consortium}). Funding for the DPAC
has been provided by national institutions, in particular the institutions
participating in the {\it Gaia} Multilateral Agreement.\end{acknowledgements}

\end{document}